%% file: main.tex
\documentclass[sigconf, review, natbib=true,anonymous=false, nonacm]{acmart}
\usepackage{adjustbox}
\usepackage{tabularx}
\usepackage{float}
\usepackage{amsmath}
\usepackage{breqn}
\usepackage{subfigure}
\usepackage{balance}

\AtBeginDocument{%
  \providecommand\BibTeX{{%
    \normalfont B\kern-0.5em{\scshape i\kern-0.25em b}\kern-0.8em\TeX}}}

\setcopyright{acmcopyright}
\copyrightyear{2018}
\acmYear{2018}
\acmDOI{XXXXXXX.XXXXXXX}

\acmConference[Conference acronym 'XX]{Make sure to enter the correct
  conference title from your rights confirmation emai}{June 03--05,
  2018}{Woodstock, NY}
\acmBooktitle{Woodstock '18: ACM Symposium on Neural Gaze Detection,
 June 03--05, 2018, Woodstock, NY} 
\acmPrice{15.00}
\acmISBN{978-1-4503-XXXX-X/18/06}

\begin{document}
\input{notation}
\fancyhead{}
\title{Unified Browsing Models for Linear and Grid Layouts}

\author{Amifa Raj}
\affiliation{%
  \institution{Boise State University}
  \city{Boise}
  \state{Idaho}
  \country{United States}
}
\email{amifaraj@u.boisestate.edu}


\author{Michael D. Ekstrand}
\affiliation{%
  \institution{Drexel University}
  \city{Philadelpha}
  \state{Pennsylvania}
  \country{United States}
}
\email{mdekstrand@drexel.edu}

\begin{abstract}

Many information access systems operationalize their results in terms of \textit{rankings}, which are then displayed to users in various ranking \textit{layouts} such as linear lists or grids.
User interaction with a retrieved item is highly dependant on the item's position in the layout, and users do not provide similar attention to every position in ranking (under any layout model).
User attention is an important component in the evaluation process of ranking, due to its use in effectiveness metrics that estimate utility as well as fairness metrics that evaluate ranking based on social and ethical concerns.
These metrics take user browsing behavior into account in their measurement strategies to estimate the attention the user is likely to provide to each item in ranking.
Research on understanding user browsing behavior has proposed several user browsing models, and further observed that user browsing behavior differs with different ranking layouts.
However, the underlying concepts of these browsing models are often similar, including varying components and parameter settings. 
We seek to leverage that similarity to represent multiple browsing models in a generalized, configurable framework which can be further extended to more complex ranking scenarios. 
In this paper, we describe a probabilistic user browsing model for linear rankings, show how they can be configured to yield models commonly used in current evaluation practice, and generalize this model to also account for browsing behaviors in grid-based layouts.
This model provides configurable framework for estimating the attention that results from user browsing activity for a range of IR evaluation and measurement applications in multiple formats, and also identifies parameters that need to be estimated through user studies to provide realistic evaluation beyond ranked lists. 
\end{abstract}

\ccsdesc[500]{Information systems~Evaluation of retrieval results}
\ccsdesc[500]{Social and professional topics~User characteristics}


\maketitle
\section{Introduction}
Information access systems (IAS) aid users in their information seeking processes by retrieving or recommending items that are estimated to be relevant to the user's information need, context, and/or preferences.
These systems often present results in ranked lists and the items are usually ordered based on their relevance score which help users to easily locate relevant items. Search engines retrieve items in response to users' explicit information need and expose them in ranked order. Unlike search engines, recommender systems identify user information preference from their interaction with the systems but similar to search engines, they also display the most relevant recommended items in ranked order. 

Various types of ranking layout are used to display the results. For example, results can be displayed in linear ranked list (figure~\ref{fig:linear ranking model}) or they can be displayed in grid view (figure~\ref{fig:grid ranking model}). In linear ranking layout, items are displayed in single-column list whereas in grid layout, items are presented in multiple rows and columns. Depending on the ranking layout, item position changes in the displayed page which affect user attention and interaction with items. Users do not provide equal attention to every item that are exposed in ranking and user attention varies based on item position in ranking \cite{yue2010beyond}. Moreover, user attention varies across ranking layouts as well \cite{xie2017investigating}. Thus user browsing behavior describing how user interact with ranking helps to estimate approximate user attention provided to items at various ranking positions.


Ranked results are often evaluated based on one or both criteria: user satisfaction such as maximum marginal relevance \cite{carbonell1998use, neurips-2020-tutorial:beyond-accuracy, moffat2008rank} and social and ethical issues such as fairness \cite{mitchell:catalog, ekstrand2022fairness, raj2022measuring, crawford17}. User browsing behavior is a significant component in evaluation metrics construction.
Evaluation metrics regarding effectiveness \cite{moffat2013users} (e.g. \textit{reciprocal rank} \cite{craswell2009mean} or \textit{nDCG}\cite{jarvelin2017ir}) and fairness (e.g. \textit{equal exposure} \cite{diaz2020evaluating, singh2018fairness} or \textit{statistical parity} \cite{sapiezynski2019quantifying}) of rankings take user browsing behavior into consideration since user attention changes with ranking position and items with similar relevance do not necessarily get the same attention in ranking. Hence, user browsing behavior is used to estimate the probability of user engagements with ranking positions which helps to measure item utility \cite{moffat2008rank} and exposure \cite{singh2018fairness} in ranking. There are research works on understanding user browsing behaviors and these studies often involve eye-tracking \cite{zhao2016gaze, djamasbi2011visual} and click models \cite{guo2020debiasing} and to date, we have multiple user browsing models. However, these browsing models work with the same underlying concept that user attention changes with ranking positions but they differ on their use of components such as, relevance information, external parameters settings. \citet{moffat2013users} identified three interchangeable functions that can be used to describe user behaviors in ranking and showed that effectiveness metrics can be represented by those functions. Their study focused particularly on linear ranking layout and generalization of effectiveness metrics for ranking.

In this work, we unify many of the extant user browsing models into a single model that accounts for both linear and grid layouts showing how particular models from the literature can be derived from specific parameterization of our model and extending them to wider range of SERP designs.
We identify the conceptual similarities among these models and disintegrate them into components and parameters level. We provide a generalized framework of user browsing models that allows researchers and practitioners to re-configure the core structure based on their required components. This structure can be further extended to more complex ranking layout scenarios.

\section{Ranking Layouts}
\begin{figure}[ht]
    \subfigure[\label{fig:lv} Linear Vertical Layout]{
        \includegraphics[width=0.28\linewidth]{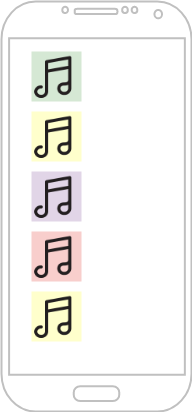}
    }
    \subfigure[\label{fig:lh} Linear Horizontal Layout]{
        \includegraphics[width=0.32\linewidth]{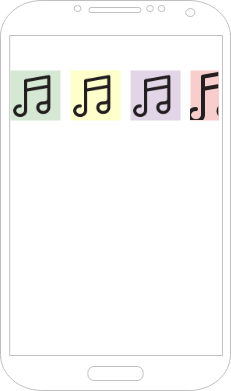}
    }
\caption{Various Types of Linear Layout Models}
\label{fig:linear ranking model}
\end{figure}

\begin{figure}[ht]
    
    \subfigure[\label{fig:wg} Wrapped Grid Layout]{
        \includegraphics[width=0.5\linewidth]{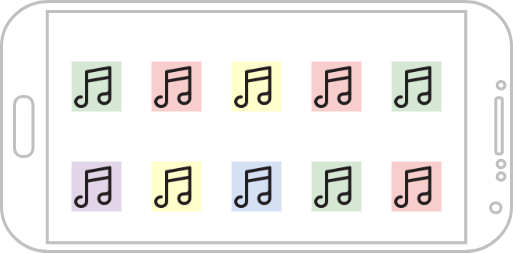}
    }
    \subfigure[\label{fig:mg} Multi-list Grid Layout]{
        \includegraphics[width=0.5\linewidth]{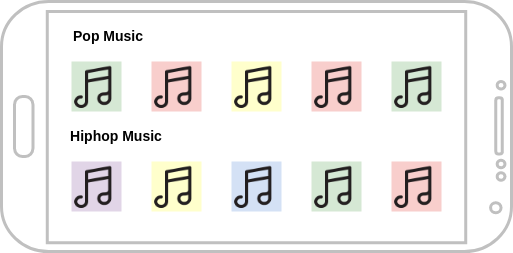}
    }

\caption{Various Types of Grid Layout Models}
\label{fig:grid ranking model}
\end{figure}

IAS display retrieved and recommended items in various ranking layouts where the items are ordered based on relevance. Based on how items are presented, we identified four types of ranking layouts and we represent the ranking layout by $r\times c$ format where $r$ is the number of rows and $c$ is the number columns. The ranking layouts can be classified into two broader categories: linear and grid layouts.
\paragraph{\textbf{Linear Layout}} Figure~\ref{fig:linear ranking model} show types of linear ranking layout.
\begin{itemize}
    \item Figures ~\ref{fig:lv} shows the \textit{vertical Linear Layout} where item are represented in a multi-row single-column format ($r\times 1$).
    \item Figure~\ref{fig:lh} shows the \textit{horizontal Linear Layout} where item are represented in a single-row multi-column format ($1\times c$).
\end{itemize}
\paragraph{\textbf{Grid Layout}} Figure~\ref{fig:grid ranking model} show types of grid ranking layout.
\begin{itemize}
    \item Figures ~\ref{fig:wg} shows the \textit{wrapped grid layout} where item are represented in a multi-row multi-column format ($r\times c$) and items are \textit{not} categorized into genres.
    \item Figure~\ref{fig:mg} shows the \textit{multi-list Linear Layout} where item are represented in a multi-row multi-column format ($r\times c$) but items are categorized into genres. Each row represents different genre or category.
\end{itemize}


\section{User Browsing Behaviors in Linear Layouts}
\label{sec:linear}
\begin{figure}[h]
\includegraphics[width=8cm]{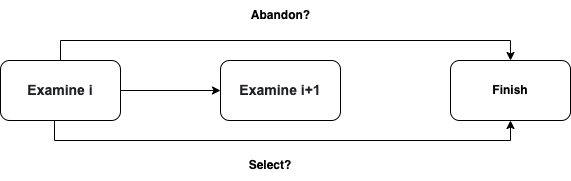}
\caption{State Transition Model of User Browsing a Linear Layout}
\label{fig:state}
\end{figure}

Several research works observed user browsing behaviors when reviewing linear ranking layouts, particularly vertically-oriented linear lists, and proposed user \textit{browsing models} to approximate the attention users are likely to pay to different items in a linear ranking. 
Two commonly-used models are the \textit{geometric} \cite{moffat2008rank} and \textit{cascade} browsing models\cite{craswell2008experimental}, each of which estimates the probability that a user will attend to the item in a particular ranking position.
These models are developed on three fundamental assumptions: 
\begin{itemize}
    \item Users browse a linear ranking from top to bottom.
    \item User attention decays with ranking positions.
    \item Users stop the process once they select an item.
\end{itemize}
Given a ranking, both geometric and cascade models approximate the probability of the user continuing to the next item. 
With similar underlying concepts, these models can be described using state transition model (figure~\ref{fig:state}). \citet{moffat2008rank} presented a similar model of user browsing behavior in linear ranked list; our model distinguishes between the \textit{selection} and \textit{abandon} events, a distinction we use in Section~\ref{sec:cascade}. Table~\ref{tab:notation} presents the list of notations used in this paper.
Given a linear ranked list $L$, for a given position $i$, user $u$ can take following actions:
\begin{enumerate}
    \item Examine ($\examine$): The user \textit{examines} (or ``visits'', ``'views'', or ``inspects'' the item at the current position ($E_i$ is the event of examining the item at position $i$).
    \item Select ($\select$): The user \textit{selects} (usually by clicking or tapping;  other work has also used the terms ``stop'' or ``click'') the item at the current position.
    \item Abandon ($\abandon$): The user terminates their browsing process without selecting an item.
    \item Continue ($C$): If the user has not select the item at the current position or abandoned the process, they move to the next position.
\end{enumerate}
\begin{table}[tb]
\centering
 \footnotesize
 \caption{Summary of notation.}
 \begin{tabular}{cl}

 \toprule
  $\doc \in \allitems$ & document or item \\
  $\user \in \allusers$ & request (user or context) \\
  $\ranking$ & ranked results of $N$ items from $\allitems$ \\
  $\iteminpos{i}$ & the item in position $i$ of linear (1-column) layout \\
  $\rankingPosition$ & rank of item $d$ in linear layout \\
  $\iteminrow{k}$ & items in $k$th row in grid layout \\
  $\iteminrowcol{k}{c}$ & items in row $k$ and column $c$ in grid layout \\
  $\relevance$ & relevance of $d$ to $q$ \\
  $\examine_i$ & event: user examines the item at position $i$ \\
  $\select_i$ & event: user selects the item at position $i$ \\
  $\abandon_i$ & event: user abandons the process after examining the item at position $i$ \\
  $\rowskip_k$ & event: user skipping the $k$th row.  \\

\bottomrule
 \end{tabular}
 \label{tab:notation}
\end{table}

The probability of continuing to the item at position $i+1$ depends on two fundamental probabilities: (1) the probability of selecting the item at position $i$ and (2) the probability of abandoning the page after examining item at position $i$.
Some treatments use other probabilities as the fundamental parameters; we discuss this more later.

\paragraph{\textbf{Selection Probability}}
In figure~\ref{fig:state}, the event ``select'' has the probability of selecting the item at position $i$; at a particular position, this probability is conditional on user \textit{examining} the item, and it can also depend on the \textit{relevance} of that item to the query. The conditional probability of user selecting item at position $i$ is:
\begin{equation}
    \label{eq:visitprob}
    P\left[\select_{i} | \examine_{i}, \relev(\ranking^{-1}(i)|\user)\right] = \stopprob(i)
\end{equation}
Hence, the \textit{selection} probability can be defined as a function of relevance:
\begin{equation}
    \stopprob(i) =
    \left\{
    \begin{array}{lr}
         \stopprob_{\mathrm{rel}} when & y=1 \\
         \stopprob_{\neg\mathrm{rel}} & y=0
    \end{array}
    \right\}
\end{equation}
When relevance is not considered, the probability of selecting the item at position $i$ is a constant: $\stopprob_{\neg\mathrm{rel}}(i) = \stopprob$.
where $\select$ is the event of selecting or clicking on an item and $\examine$ refers to the event of examining or viewing that item. 
This selection probability can be modified based on the availability and the type of relevance information.

\paragraph{\textbf{Abandon Probability}}
User can abandon the page with \textit{abandon} probability $\abandonprob$ which can be fixed at each position that has the effect of being cumulative over the ranking positions. The \textit{abandon} probability can be derived as position-based exponential decay and it can also be a extended as a conditional probability function which can be dependant on relevance of items.

Therefore, the continuation probability or the probability of moving to the next position can be derived from selection probability and abandon probability of the current position. 
Users will continue to the item at position $i+1$ if they have not selected the item at position $i$ and did not abandon the process after examining the item and position $i$.
For a given linear ranking $\ranking$, the generalized model for predicting the probability of user examining the item $d$ at a particular position $i$ when user attention decays exponentially is:
\begin{equation}
    \label{eq:general_linear}
    P[E_i] = (1-\abandonprob)^{i-1}\prod_{j\in[1,i-1)}(1-\stopprob(j))
\end{equation}

\subsection{Static User Browsing Models}
In static user browsing models, only item position is taken into account to estimate user attention to items in ranking \cite{moffat2008rank, chapelle2009expected}.
\citet{moffat2008rank} considered \textit{geometric} browsing model to propose \textit{rank-biased precision} metric (\textit{RBP)} which is an effectiveness metric for linear ranking. In their assumed browsing model, user attention decays exponentially with ranking positions and the probability of user continuing to the next positions depends on the probability of user selecting item at the current positions. They proposed a \textit{persistence} or \textit{continuation} probability $\contprob$ to derive the possibility of user continuing to the next position. In this model, the \textit{continuation} probability is not dependant on relevance of items and it is considered as a constant, hence, their proposed model is not distinguishing between \textit{selection} probability and \textit{abandon} probability.
In this  model, user will always examine item at the first position and hence the probability of examining item at $i$th position is $\contprob^{i-1}$.
For a given ranking $\ranking$ of size $N$, \textit{rank-biased precision} can be derived as 
\begin{align*}
    \operatorname{RBP}(\ranking) = (1-\contprob) \sum_{i=1,2,...N}y\left(\iteminpos{i}\middle|q\right)\contprob^{i-1}
\end{align*}
where $y\left(\iteminpos{i}\middle|q\right)$ denotes the relevance of item in position $i$ to the user request $q$. 
Since the \textit{persistence} probability $\contprob$ was not dependant on relevance in geometric user browsing model, the probability of examining item $d$ at ranking position $i$ is:
\begin{align*}
    P_{\mathrm{geometric}}[E_i] = \prod_{i\in[1,i-1)} \contprob
\end{align*}
which can be derived from equation~\ref{eq:visitprob} without considering relevance information and \textit{abandon} probability. 
\begin{equation}
    \label{eq:geom}
    P[E_i] = \prod_{j\in[1,i-1)}(1-\stopprob_{\neg\mathrm{rel}})
\end{equation}
\citet{biega2018equity} proposed another version of \textit{geometric} model where the attention decays geometrically and each position has the equal probability of being selected.
Whether we want to consider examine probability of prior positions or not, we can represent both versions of \textit{geometric} model through equation~\ref{eq:visitprob}.

\subsection{Cascade Models}
\label{sec:cascade}
\citet{craswell2008experimental} proposed another user browsing model that incorporates item relevance when estimating users' item selection behavior. Their proposed \textit{cascade} click model (called \textit{adaptive} by \citep{moffat2013users}) incorporates item relevance into the selection process, so that users are much more likely to select a relevant document than an irrelevant one; the probability of examining an item at a particular ranking position therefore depends on the relevance of items in previous positions.
Specifically, the probability of user clicking or selecting an item $d$ is a function of $y(d|q)$ (the relevance of $d$ for a given query $q$, which may be binary or graded). 
The event of user selecting an item at position $i$ depends on the probability of user selecting an item at position $i$ and users skipped (did not select) all the items prior to that position which are dependant on the relevance of items.
Hence, the the probability of user clicking or selecting an item at position $i$ can be derived from the fundamental relevance-dependant selection probability $\stopprob_{\mathrm{rel}}$:
\begin{align*}
    P[S_i] = \stopprob_{\mathrm{rel}}(i) \prod_{j\in[1, i-1)}(1-\stopprob_{\mathrm{rel}}(j)) 
\end{align*}

In both the cascade and geometric models, user attention decays exponentially with ranking position, but in the cascade model the user is much more likely to stop at a relevant item, so the examination and selection probabilities at a particular position differ from ranking to ranking, and have has jumps at the positions of relevant items.
\begin{equation}
    \label{eq:cas}
    P_{\mathrm{cascade}}[E_i] = \prod_{j\in[1,i-1)} (1-\stopprob_{\mathrm{rel}}(j))
\end{equation}

where $\stopprob_{\mathrm{rel}}(j)$ depends on the relevance of the item in position $j$ to the user request $q$ and the relevance can be binary or graded. 

\citet{chapelle2009expected} use a cascade model to derive an effectiveness metric, \textit{expected reciprocal rank} (ERR).
Unlike \textit{RBP} metric, \textit{ERR} depends on the relevance of items to infer probability of user selecting an item at particular ranking position and directly derives effectiveness from those probabilities instead of using them to weight a precision metric.
For a given ranking of size $N$,
\begin{align*}
    ERR(\ranking) = \sum_{i=1,2..N} \frac{1}{i}P[S_i]
\end{align*}
where $P[S_i]$ is the probability of clicking or selecting an item at position $i$ which is same as the \textit{cascade} click model.

\citeauthor{chapelle2009expected} extended the cascade model through an additional parameter, an \textit{abandonment} probability modeling the probability of user terminating their browsing regardless of whether they have selected an item (either to abandon the search entirely, or to reformulate their query).
In this extended \textit{cascade} model, the probability of an user examining an item at position $i$ is:
\begin{equation}
    P[E_i] = (1-\abandonprob)^{i-1}\prod_{j\in[1,i-1)}(1-\stopprob_{\mathrm{rel}}(j)) 
\end{equation}

Therefore, the probability of examining the item at position $i$ can be derived using equation~\ref{eq:general_linear} incorporating abandon probability and relevance information.
\begin{table*}[]
\caption{Parameters of Browsing Models and the range of parameter values}
\label{tab:parameters}
\begin{minipage}{\textwidth}
\centering\small
\begin{tabular}{llll}
\hline
\textbf{Parameters} & \textbf{Name}            & \textbf{Description}                               & \textbf{Values}        \\ \hline
$\stopprob$         & Selection Probability    & Probability of selecting an item at position $i$   & \{0.1, 0.2, ..., 0.9\} \\
$\abandonprob$       & Abandon Probability      & Probability of abandoning the process.             & \{0.1, 0.2, ..., 0.9\} \\
$\contprob$         & Continuation Probability & Probability of continuing to the position $i$      & \{0.1, 0.2, ..., 0.9\} \\
$\skipprob$         & Skipping Probability     & Probability of skipping an entire row.             & \{0.1, 0.2, ..., 0.9\} \\
$\slowdecay$        & Decay                    & Incorporate slow browsing tendency for grid layout & \{1.1, 1.2, ..., 2.0\} \\ \hline
\end{tabular}
\end{minipage}
\end{table*}

\subsection{Unifying Ranking Browsing Models}

We can see that the geometric and cascade browsing models are capturing the same fundamental ideas, with the difference that the cascade model incorporates item relevance into \textit{selection} probabilities.
We can therefore derive extant browsing models from our state model with two main probability parameters: the \textit{selection} probability and the \textit{abandon} probability, where the \textit{selection} probability may or may not depend on item relevance. \textit{Abandon} probability can also be extended as a conditional probability of relevance.
Our generalized browsing model can be configured to implement various browsing models in the following ways:
\begin{itemize}
    \item To use geometric browsing model, the relevance component of \textit{selection} probability function and the \textit{abandon} probability will be ignored (setting the value as $\abandonprob = 0$).
    \item To use cascade browsing model, the relevance component of \textit{selection} probability can be binary or graded. 
    \item Both models can further incorporate \textit{abandon} probability by setting appropriate parameter values.
\end{itemize}
Table~\ref{tab:parameters} shows the list of discussed parameters that can be incorporated in browsing models and their range of acceptable values.  
With suitable parameter choices, our generalized model can therefore realize a wide range of probabilistic attention models both from the literature and yet to be devised.
 
\section{Extending Generalized Framework to Grid Layout}
In this section, we will further extend the generalized user browsing model for linear ranking layout to grid layout. 

\begin{figure}[]
\includegraphics[width=8cm]{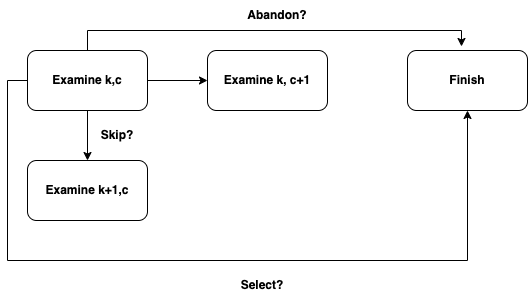}
\caption{State Transition Model of User Browsing a Grid Layout}
\label{fig:gridstate}
\end{figure}

\subsection{Linear Layout is Single-Column Grid Layout}
In grid-based layout, users have similar actions as linear ranking layout with an additional possibility of \textit{skipping row} action. Unlike linear ranking layout, in grid layouts, users can skip an entire row and move to the next row. We denote this \textit{row-skipping} event as $\rowskip$. User actions in grid-based ranking can be demonstrated by figure~\ref{fig:gridstate}, the state-transition model. Therefore, for a given grid-based ranking $\ranking$ the probability of examining item at position $\iteminrowcol{k}{c+1}$ (row is $k$ and the column is $c+1$) depends on:
\begin{itemize}
    \item the probability of selecting the item at position $\iteminrowcol{k}{c}$,
    \item the probability of abandoning the process after examining item at position $\iteminrowcol{k}{c}$, and
    \item the probability of skipping row $k$ after examining item at position $\iteminrowcol{k}{c}$.
\end{itemize} 
For 1-column grid layout or linear vertical layout, the row skipping probability can be ignored. 
Hence, the continuation probability of or the probability of moving to the next position in grid layout can be derived from \textit{selection} probability, \textit{row-skipping} probability, and \textit{abandon} probability at the current position. The generalized conditional probability of user selecting the item position $i$ in grid ranking $\ranking$ is:
\begin{equation}
    \label{eq:visitprobgrid}
    P\left[\select_{\iteminrowcol{k}{c}} | \examine_{\iteminrowcol{k}{c}}, \relev\left(\iteminrowcol{k}{c}|\user\right)\right] = \stopprob(\iteminrowcol{k}{c})
\end{equation}

\subsection{User Browsing Models for Grid Layouts}
Several studies have sought to understand user browsing behavior in grid layouts, identifying that user browsing behavior in a grid layout is different than it is in linear layouts. Users do not necessarily view top-to-bottom while interacting with a grid-based interface; rather, they show various distinct tendencies such as \textit{central fixation} \cite{tatler2007central} and \textit{F-shaped browsing} \cite{djamasbi2011visual}. \citet{tatler2007central} did an user eye-movement study and showed that users have the tendency of \textit{central-fixation} while viewing images showing on screen and that tendency can persist for grid-based image search results. Eye-tracking studies on grid-based web-pages \cite{djamasbi2011visual} and on grid-based recommendation \cite{zhao2016gaze} show that users often follow \textit{F-shaped} viewing pattern but both studies indicate that user browsing behavior depends on task and content. 
In grid-based image search results, \citet{xie2017investigating, xie2019grid} observed several user browsing behavior tendencies through eye-tracking study; (1) \textit{slower-decay} where user attention decays at a slower rate in grid layout than in linear layout, (2) \textit{row-skipping} where users skip row and moves to the next row, and (3) \textit{middle-bias} where users put more attention to items at the middle positions.
However, \citet{shrestha2007eye, balyan2021using} highlighted to the need of considering page content while studying user behavior for web-pages and item meta information for e-commerce grid-based product search results. 

\subsection{Generalized Browsing Model}

Linear vertical ranking layouts can be treated as 1-column grid layouts, and the extant browsing models for linear vertical ranking can be further modified for grid layouts (and still capture the linear behavior when the number of columns is set to 1).
\citet{guo2020debiasing} observed the similar user browsing pattern as \citet{xie2019grid} in grid-based e-commerce products search results and they proposed modified \textit{geometric} (equation~\ref{eq:geom})\footnote{The original study referred the model as cascade click model. However, in our paper, we referred the model as geometric to keep the conceptual consistency.} browsing models incorporating \textit{slower-decay} and \textit{row-skipping} grid-based layout specific user browsing behaviors to generate user attention model suitable for grid-based e-commerce search results.

For \textit{slower-decay}, a \textit{decay} parameter $\slowdecay$ is used to incorporate users slow browsing patterns for grid-based layout. These parameters can be plugged into any linear browsing model to make the browsing model suitable for the grid layout. The visiting probability of item $d$ at ranking position $i$ where the row number is $r(i)$ and column is $c(i)$  which is:
\begin{equation}
    P_{SD}[E_i] = \operatorname{min}(\slowdecay^{r(i)} \prod_{j=[1, i-1]}(1-\stopprob(j), 1))
\end{equation}
The selection probability may or may not incorporate relevance judgements or estimates. The parameter $\slowdecay$ can be modeled by \textit{abandon} probability with the assumption of the probability of abandoning items increases with rows; as users move vertically in grid-layout, the probability of abandoning the process increases. Hence, the \textit{slower-decay} user browsing behavior can be modeled by adjusting \textit{abandon} probability where the \textit{abandon} probability will increase vertically with rows but the horizontal (columns) \textit{abandon} probability will remain the same across the columns in each row. 

\begin{equation}
    P_{SD}[E_i] = \abandonprob^{r(i)-1} \prod_{j=[1, i-1]}(1-\stopprob(j), 1))
\end{equation}

Users' \textit{row-skipping} behavior is incorporated in browsing model with an assumption that if users \textit{examine} any item in a row, that particular row is not \textit{skipped}. The parameter $\skipprob$ determines the probability of \textit{skipping} each of the rows before $r(i)$. If users \textit{examined} or \textit{selected} any item in a particular row, that means that row was not skipped. Hence the examining probability of item $d$ at ranking position $i$ (row $r(i)$, column $c(i)$) in the generalized browsing model is:


\begin{equation}
    P_{RS}[E_i] = \left[\prod_{k=1}^{r(i)-1}(1-\skipprob) \prod_{j\in \iteminrow{k}} (1-\stopprob(j)) + \prod_{k=1}^{r(i)-1}\skipprob\right] \prod_{i\in \iteminrow{r(i)}}(1-\stopprob(i))
\end{equation}

Therefore, if we want to incorporate all the components and browsing behaviors, the generalized browsing model of estimating the probability of user examining item at position $i$ in a given ranking is:

\begin{equation}
    \label{eq:general}
    \begin{split}
        P[E_i] = & P[\neg\abandon_{r(i)}] \left[\prod_{k=1}^{r(i)-1}P[\neg\rowskip_{k}]\prod_{j\in \iteminrow{k}}P[\neg\abandon_j|\neg\select_j, \examine_j])) + \prod_{k=1}^{r(i)-1}P[\rowskip_{k}]\right] \\
    & \prod_{i\in \iteminrow{r(i)}}P[\neg\abandon_i|\neg\select_i, \examine_i])
    \end{split}
\end{equation}


Our generalized browsing model can be configured to implement various browsing models and ranking layouts in the following ways:
\begin{itemize}
    \item To consider \textit{slower-decay} user browsing behavior without \textit{row-skipping} tendency, the \textit{row-skipping} probability $\skipprob$ can be ignored with the value of 0.
    \item To consider a linear vertical layout or single-column ranking, the generalized model can be used by ignoring \textit{row-skipping} and \textit{slower-decay} behavior. In that case, the probability of skipping a row will be fixed ($\skipprob=0$) and the column abandon probability will be 0 with a constant row abandon probability.
    \item The configuration of the discussed grid-layout suitable browsing models depend on the parameterization of \textit{selection} probability, \textit{row-skipping} probability,  and \textit{abandon} probability. Table~\ref{tab:parameters} show the potential values of the mentioned parameters.
    \item Based on the availability of relevance, the \textit{selection} probability $\stopprob(i)$ can be relevance dependant (\textit{cascade}) or constant (\textit{geometric}).
    \item The browsing models can incorporate \textit{abandon} probability $\abandonprob$ at each position of the ranking.
    \item The abandon probability can be derived as a function of relevance of items.
\end{itemize}

This generalized browsing model can be further extended to incorporate various browsing patterns and complex ranking layouts. 
\begin{itemize}
    \item The generalized browsing model can be extended to incorporate \textit{middle-bias} which is a grid layout suitable user browsing behavior by increasing \textit{selection} probability for the middle positions in ranking. \citet{xie2019grid} modified \textit{selection} probability by considering that as a normal distribution. 
    \item This generalized browsing model can also incorporate \textit{F-shaped} user browsing tendency by adjusting \textit{skipping} probability $\skipprob$ and \textit{selection} probability $\stopprob$. However, we need accurate user browsing pattern to derive the appropriate value of the parameters.
\end{itemize}
\section{Conclusion and Future Research Direction}
In this work, we identify user browsing models for linear ranking layout in information access systems and show that the extant user browsing models are conceptually similar and they can be generalized and configured based on ranking layouts, availability of component like relevance information, and parameter settings. We provide generalized configurable framework of the browsing models that can be extended to grid layout by considering grid-layout suitable browsing behaviors. The proposed unified framework relies on configurable parameters such as \textit{selection} probability, \textit{abandon} probability, \textit{row-skipping} probability, and \textit{decay} parameters and thus various browsing behavior in various ranking layouts can be represented by calibrating these parameters. 

Our analysis indicates the importance of knowing accurate user browsing behaviors for various ranking layouts to estimate optimal parameter values and user-eye-tracking studies in various ranking scenarios can help in this area. 
This work relies on multiple existing assumptions of user browsing behaviors (mentioned in section~\ref{sec:linear}) excluding other possible user browsing behaviors that are common but seldom studied. For example, we assumed that the process will end once user selects an item. However, users may select an item and return to the result page. Future user studies can consider this browsing behavior so that browsing models can incorporate this \textit{multi-select} user behavior; our theoretical treatment can be extended to account for it by adding an additional probability modeling whether the user continues or stops their browsing after selecting an item.

Furthermore, in grid layout, users can skip a row even after examining some items in that particular row and examining items in a row may not always follow left-to-right pattern. Future research work on understanding user browsing behaviors in grid layout can focus on identifying users' row-skipping behavior and their patterns in examining items in rows. Then the generalized browsing models can be further modified and configured depending on users' browsing patterns in grid-layout. Therefore, studies on user browsing behavior can emphasize on inferring optimal parameter setting to generate reliable user attention scores for any given ranking layouts. Users' browsing behavior in \textit{multi-list} grid-layout is still under-studied and the categories or genres can have affect on users browsing behavior. Hence, \textit{wrapped} grid-layout suitable browsing models may not be applicable in \textit{multi-list} grid layout which indicate the need to understanding user browsing behaviors in \textit{multi-list} ranking scenarios.





\bibliographystyle{ACM-Reference-Format}
\bibliography{reference}

\end{document}

%% file: notation.tex
\newcommand{\ranking}[0]{L}
\newcommand{\allitems}[0]{D}
\newcommand{\allusers}[0]{Q}

\newcommand{\doc}[0]{d}
\newcommand{\docs}[1]{\doc_{#1}}
\newcommand{\user}[0]{q}
\newcommand{\users}[1]{q_{#1}}

\newcommand{\prefix}[1]{\ranking_{\le {#1}}}

\newcommand{\rankingPosition}[0]{\ranking(\doc)}
\newcommand{\iteminpos}[1]{\ranking_{\mathrm{#1}}}
\newcommand{\relev}[0]{y}
\newcommand{\relevance}[0]{y(d|q)}
\newcommand{\predictedrelevance}[0]{\hat{y}(d|q)}

\newcommand{\group}[0]{g}
\newcommand{\groups}[0]{\mathcal{G}}
\newcommand{\groupname}[1]{\groups^{#1}}

\newcommand{\progroup}[0]{\groups^+}
\newcommand{\nonprogroup}[0]{\groups^-}

\newcommand{\Protected}[0]{\progroup(\ranking)}
\newcommand{\nonProtected}[0]{\nonprogroup(\ranking)}

\newcommand{\alignmentvec}[0]{\groups(\doc)}
\newcommand{\alignmentmat}[0]{\groups(\ranking)}

\newcommand{\populationEstimator}[0]{\hat{\textbf{p}}}
\newcommand{\probabiltyDist}[0]{\hat p}

\newcommand{\attention}[0]{\mathbf{a}}
\newcommand{\attentionvec}[0]{\attention_{\ranking}(\doc)}
\newcommand{\attentionmat}[0]{\attention_{\ranking}}
\newcommand{\exposure}[0]{\boldsymbol\epsilon}
\newcommand{\groupexposure}[0]{\exposure_{\ranking}}

\newcommand{\stochasticRanking}[0]{\pi}
\newcommand{\targetPolicy}{\tau}

\newcommand{\prefd}[0]{\mathrm{PreF}_\Delta}
\newcommand{\AWRF}[0]{\mathrm{AWRF}_\Delta}

\newcommand{\expectation}[2]{\operatorname{E}_{#1}[#2]}
\newcommand{\contprob}[0]{\lambda}
\newcommand{\stopprob}[0]{\psi}
\newcommand{\abandonprob}[0]{\alpha}
\newcommand{\slowdecay}[0]{\beta}
\newcommand{\skipprob}[0]{\gamma}

\newcommand{\row}[0]{{\mathrm{row}}}
\newcommand{\iteminrow}[1]{\ranking_{\mathrm{#1}, \cdot}}
\newcommand{\iteminrowcol}[2]{\ranking_{\mathrm{#1}, \mathrm{#2}}}
\newcommand{\itemrow}[0]{\row(\doc)}

\newcommand{\select}[0]{\textit{S}}
\newcommand{\examine}[0]{\textit{E}}
\newcommand{\abandon}[0]{\textit{A}}
\newcommand{\rowskip}[0]{\textit{K}}